# Dynamics of Li-ion in $V_2O_5$ Layers from First-Principles Calculations

Baltej Singh[1,2], M. K. Gupta[1], R. Mittal[1,2], and S. L. Chaplot[1,2]
[1]*Solid State Physics Division, Bhabha Atomic Research Centre, Mumbai, 400085, India*
[2]*Homi Bhabha National Institute, Anushaktinagar, Mumbai 400094, India*

The alkali atoms, due to their small sizes and low charge ionic states, are most eligible to intercalate in the structural layers of $V_2O_5$. We have applied *ab-initio* density functional theory to study the dynamics of Li-ion in layers of α-$V_2O_5$. The calculations are performed for two compositions, namely, $Li_{0.08}V_2O_5$ and $Li_{0.16}V_2O_5$, and show that there are unstable phonon frequencies. The unstable modes have large amplitude of Li atom along the *b*-axis of the orthorhombic unit cell indicating that such unstable modes could initiate Li-ion diffusion along *b*-axis. The *ab-initio* molecular dynamics simulations are performed up to 25 ps at 1200 K, which reveal one-dimensional diffusion of Li atoms. The diffusion pathways of Li atoms from the simulations seem to follow the eigenvectors of the unstable phonon modes obtained in the intercalated structure.

# I. Introduction

The battery research is progressively increasing for efficient storage of energy. This demands the batteries to be lighter, quickly chargeable, safer, long lived, stable and cost effective[1, 2]. There are relatively few gains that can be made to the battery besides directly improving component materials. In this respect many material scientists are searching for efficient electrodes[3-5] and electrolyte materials[6-10]. It has been calculated[11] that an increase of 57% in cell energy density can be obtained by doubling the capacity of cathode; however, only 47% gain is obtained by 10 times increasing the capacity of anode. Therefore, large focus of electrode material research is dedicated to the discovery of cathode materials[3, 4, 12].

Layered structure materials have been suggested best for battery cathode purpose[13, 14]. The present battery technology uses layered lithium cobalt oxide ($LiCoO_2$)[15] and various doped combinations like NMC ($LiNi_{0.33}Co_{0.33}Mn_{0.33}O_2$), NCA ($LiNi_{0.8}Co_{0.15}Al_{0.05}O_2$) etc. as a cathode[16]. The major problem with these materials is that they contain cobalt in significant amount and hence limited by toxicity, high cost, low thermal stability, and fast capacity fade at high current rates. Therefore, many other alternative materials are being studied for this purpose[1, 3, 13, 14, 16]. The low cost, high abundance, easy synthesis and high energy density [17]of vanadium pentoxide ($V_2O_5$) makes it useful as a cathode material for Li, Na and Mg ion batteries [5, 18-22].

The compound $V_2O_5$, due to its layered structure, offers a large affinity to host and intercalate with small elements [23-25]. The alkali atoms, due to their smaller sizes and lower charge ionic states, are most eligible to intercalate in the structural layers of $V_2O_5$. The lower charge states of these alkali ions favour their reversible intercalation. This type of intercalation is very useful for battery technology, where electrode materials are required to intercalate and deintercalate Li ions for charging and discharging of Li ion battery. Therefore, layered structures with high stability have always been of interest to battery researchers. In this respect, a few well known layered structures like $MoSe_2$, $WSe_2$, $MoS_2$, $WS_2$ etc have been studied for their thermodynamical behaviour for intercalation with Li, Na and Mg ions[26-28]. The intercalation of alkali ions like Li, Na and Mg has also been studied in various phases of $V_2O_5$ using *ab-initio* DFT techniques[24, 25]. The structures are found to form some metastable phases upon intercalations. Multivalent ions inserted within these structures encounter suboptimal coordination environments and expanded transition states, which facilitate easier ion diffusion. The calculated activation energy barrier using nudged elastic band (NEB) method for divalent Mg ion is found in the range of 0.2 to 0.8 eV in various phases [24]. This barrier reduces to 0.11−0.16 eV for Li diffusion in α-$V_2O_5$.

In the present study, we have performed the ab-initio molecular dynamics simulations to understand the dynamics of Li-ions among the $V_2O_5$ layers at high temperature. The MD trajectories are used to obtain the mechanism of Li diffusion. Moreover, the density functional perturbation theory (DFPT) is used to calculate the zone centre vibrational frequencies of Li intercalated $V_2O_5$ which are further used to investigate role of phonons in Li diffusion.

## II. Calculations

The calculations are performed using Vienna ab-initio Simulation Package [29, 30] (VASP). All the simulations are performed on *1×3×2* supercell of the orthorhombic α-$V_2O_5$ phase. The kinetic energy cutoff of 960 eV is adapted for plane wave pseudo-potential. A k-point sampling with a grid of 2×2×2, generated automatically using the Monkhorst-Pack method[31], is used for structure optimizations. The ab initio molecular dynamics simulations are performed in NVT ensemble. Newton's equations of motion are integrated using Verlet's algorithm with a time step of 1 fs.

## III. Results and Discussions

A supercell, *1×3×2,* of the ambient α-$V_2O_5$ containing 84 atoms, is intercalated with one and two Li atoms giving rise to $Li_{0.08}V_2O_5$ and $Li_{0.16}V_2O_5$ configurations, respectively. The Brillouin zone centre phonon frequencies for these two supercell structures are calculated using linear response density functional perturbation approach. The calculated range of phonon frequencies is nearly same in both the structures (Fig. 1). These calculations show that there are one and two unstable phonon frequencies in the supercells of $Li_{0.08}V_2O_5$ and $Li_{0.16}V_2O_5$, respectively. This means that there is one unstable mode per Li-atom. The eigenvectors of these phonon modes are shown in Fig 2. The unstable mode in $Li_{0.08}V_2O_5$ has large amplitude of Li atom along *b*-axis of the orthorhombic unit cell. The large magnitude arises from the open channels along *b*-axis in the layers of $V_2O_5$ structure. On the other hand, in case of $Li_{0.16}V_2O_5$, calculations give two unstable phonon frequencies. The eigenvectors corresponding to these frequencies give two different displacement patterns of Li atoms movements along *b*-axis. In one case, both the Li atoms move opposite to each other (Fig 2), while in the second unstable phonon mode, both the Li atoms move in the same direction along the *b*-axis. These unstable modes may be responsible for Li diffusion in the structure along *b*-axis. However, molecular dynamics simulations are required to fully understand the nature and pathways of Li diffusion in $V_2O_5$.

To investigate the Li diffusion in Li intercalated $V_2O_5$ we have performed the *ab-initio* molecular dynamics simulations up to about 25 ps at 1200 K. The calculated squared displacements (SD) of various atoms in $Li_{0.08}V_2O_5$ and $Li_{0.16}V_2O_5$ as obtained from these simulations are shown in Fig 3 and Fig 4 respectively. It can be seen from Fig 3 (right) that Li atoms show large jumps in SD value with time while the host structure atoms (V, O) have small SD and do not show any time dependence implying stability of host $V_2O_5$ structure. Further, the calculated anisotropic SD analysis shows that the Li jumps are taking place along the b- axis. There are no jumps along the c-direction implying the absence of interlayer Li diffusion in $V_2O_5$. This is in accordance with the calculated eigenvector pattern (Fig. 2) of unstable zone centre phonon mode in $Li_{0.08}V_2O_5$.

The simulations for $Li_{0.16}V_2O_5$ at 1200 K performed for 25 ps show similar rigidity of V-O structure like $Li_{0.08}V_2O_5$ (Fig 4). Both the Li atoms present in the succeeding layers of $V_2O_5$ show jump like diffusion. The calculated anisotropic displacements show that the jumps of both Li atoms are only along the b- direction in the unit cell. These jumps show interesting behaviour that when one Li atom has maximum amplitude, another li-atom in second layer has low amplitude. This might be due to the fact that the diffusion of one Li atom locally distorts the structure, which may create hindrance for the other Li in the second layer. The Coulombic repulsion between the two Li atoms in different layers would also act as repulsive push for Li diffusion. The direction of Li jumps is consistent with the calculated eigenvector pattern of unstable zone centre phonon modes in $Li_{0.16}V_2O_5$. We have observed that during the jump from one site to another when the Li approaches towards the saddle point, i. e., where it faces the direct interaction with oxygen, it gets the largest push for jump.

The other interesting point to note here is the difference in the magnitude of the SD of Li in case of $Li_{0.08}V_2O_5$ and $Li_{0.16}V_2O_5$. The large values of SD of up to about 200 $Å^2$ in about 25 ps in $Li_{0.16}V_2O_5$ in comparison to that of 60 $Å^2$ in $Li_{0.08}V_2O_5$ may arise from the fact that there is an increase of about 3 % in the area of the channel like structure in the a-b plane with structure of two Li atoms (a=11.694 Å, b=10.721 Å, c=9.886 Å) in comparison to that with the structure with single Li atom (a=11.483 Å, b=10.634 Å, c=9.949 Å). This would facilitate the increase in diffusion. However, the separation between the two channels along the c-axis reduces slightly by about 0.6 % for the structure with two lithium atoms.

The calculated trajectories of Li in $Li_{0.08}V_2O_5$ and $Li_{0.16}V_2O_5$ obtained from ab-initio MD simulations at 1200 K are shown in Fig. 5. The one-dimensional diffusion along the b-axis is clearly evident. Overall, both the unstable phonon modes and the actual diffusion pathways from MD simulations show diffusion of intercalated Li along the *b*-axis of the structure. The diffusion of Li seems to follow the same path as guided by the eigenvector of unstable phonon modes in $Li_xV_2O_5$. The vibrational dynamics may provide initial screening for diffusion pathways for Li ion migration in solids.

## IV. Conclusions

We have investigated the dynamics of Li ions in $V_2O_5$ layers, using ab-initio molecular dynamics simulations. Moreover, the phonon frequencies and corresponding eigenvectors in the intercalated $V_2O_5$ are calculated using lattice dynamics to see the role of vibrational dynamics in the onset of Li diffusion in the compound. The combined *ab-initio* molecular dynamical simulation and lattice dynamics study shows that the vibrational dynamics of Li may initiate one-dimensional diffusion of Li along the *b*-axis in $Li_xV_2O_5$.

[14] Q. Cheng, W. He, X. Zhang, M. Li, L. Wang, Modification of Li2MnSiO4 cathode materials for lithium-ion batteries: a review, Journal of Materials Chemistry A, 5 (2017) 10772-10797.

[15] K. Mizushima, P. Jones, P. Wiseman, J.B. Goodenough, LixCoO2 (0< x<-1): A new cathode material for batteries of high energy density, Materials Research Bulletin, 15 (1980) 783-789.

[16] B. Xu, D. Qian, Z. Wang, Y.S. Meng, Recent progress in cathode materials research for advanced lithium ion batteries, Materials Science and Engineering: R: Reports, 73 (2012) 51-65.

[17] M. Sathiya, A.S. Prakash, K. Ramesha, J.M. Tarascon, A.K. Shukla, V2O5-Anchored Carbon Nanotubes for Enhanced Electrochemical Energy Storage, Journal of the American Chemical Society, 133 (2011) 16291-16299.

[18] D. Yu, C. Chen, S. Xie, Y. Liu, K. Park, X. Zhou, Q. Zhang, J. Li, G. Cao, Mesoporous vanadium pentoxide nanofibers with significantly enhanced Li-ion storage properties by electrospinning, Energy & Environmental Science, 4 (2011) 858-861.

[19] Q. Liu, Z.-F. Li, Y. Liu, H. Zhang, Y. Ren, C.-J. Sun, W. Lu, Y. Zhou, L. Stanciu, E.A. Stach, J. Xie, Graphene-modified nanostructured vanadium pentoxide hybrids with extraordinary electrochemical performance for Li-ion batteries, Nat Commun, 6 (2015).

[20] Q. Wei, J. Liu, W. Feng, J. Sheng, X. Tian, L. He, Q. An, L. Mai, Hydrated vanadium pentoxide with superior sodium storage capacity, Journal of Materials Chemistry A, 3 (2015) 8070-8075.

[21] Q. An, Q. Wei, P. Zhang, J. Sheng, K.M. Hercule, F. Lv, Q. Wang, X. Wei, L. Mai, Three-Dimensional Interconnected Vanadium Pentoxide Nanonetwork Cathode for High-Rate Long-Life Lithium Batteries, Small, 11 (2015) 2654-2660.

[22] M. Rao, Vanadium Pentoxide Cathode Material for Fabrication of all Solid State Lithium-Ion Batteries-a Case Study, Research Journal of Recent Sciences, 2 (2013) 67-73.

[23] A. Parija, Y. Liang, J.L. Andrews, L.R. De Jesus, D. Prendergast, S. Banerjee, Topochemically de-intercalated phases of V2O5 as cathode materials for multivalent intercalation batteries: a first-principles evaluation, Chemistry of Materials, 28 (2016) 5611-5620.

[24] A. Mukherjee, N. Sa, P.J. Phillips, A. Burrell, J. Vaughey, R.F. Klie, Direct investigation of Mg intercalation into the orthorhombic V2O5 cathode using atomic-resolution transmission electron microscopy, Chemistry of Materials, 29 (2017) 2218-2226.

[25] G. Sai Gautam, P. Canepa, A. Abdellahi, A. Urban, R. Malik, G. Ceder, The intercalation phase diagram of Mg in V2O5 from first-principles, Chemistry of Materials, 27 (2015) 3733-3742.

[26] X. Wang, X. Shen, Z. Wang, R. Yu, L. Chen, Atomic-Scale Clarification of Structural Transition of MoS2 upon Sodium Intercalation, ACS Nano, 8 (2014) 11394-11400.7

**FIG. 1** The calculated zone centre optic phonon modes in a *1×3×2* supercell of $Li_{0.08}V_2O_5$ and $Li_{0.16}V_2O_5$. The vertical lines in the plot corresponds to the calculated frequencies of zone centre optic phonon modes.

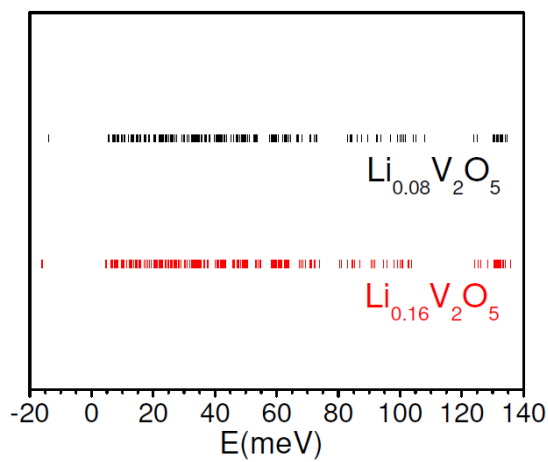

**FIG 2** The calculated eigenvector of unstable phonon modes in **(a)** $Li_{0.08}V_2O_5$ and **(b)** $Li_{0.16}V_2O_5$ obtained from density functional perturbation theory approach. Key: O- Red & V-Blue. The Li atoms are shown by green circles. The arrow on the green balls gives the direction of displacement of Li atoms.

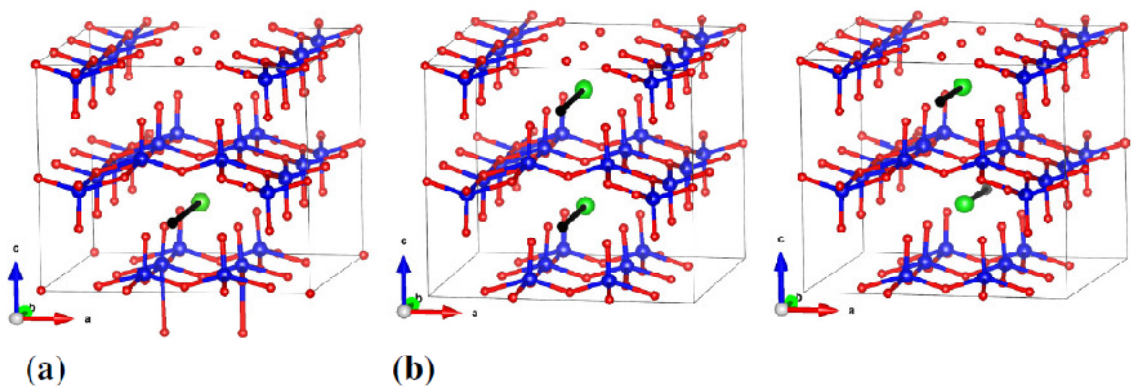

**FIG 3** (Left) The calculated anisotropic squared displacement of Li atom along the crystallographic axes at 1200 K. (Right) The calculated mean squared displacement of various atoms Li, O, Al in $Li_{0.08}V_2O_5$ obtained from ab-initio MD simulations at 1200 K.

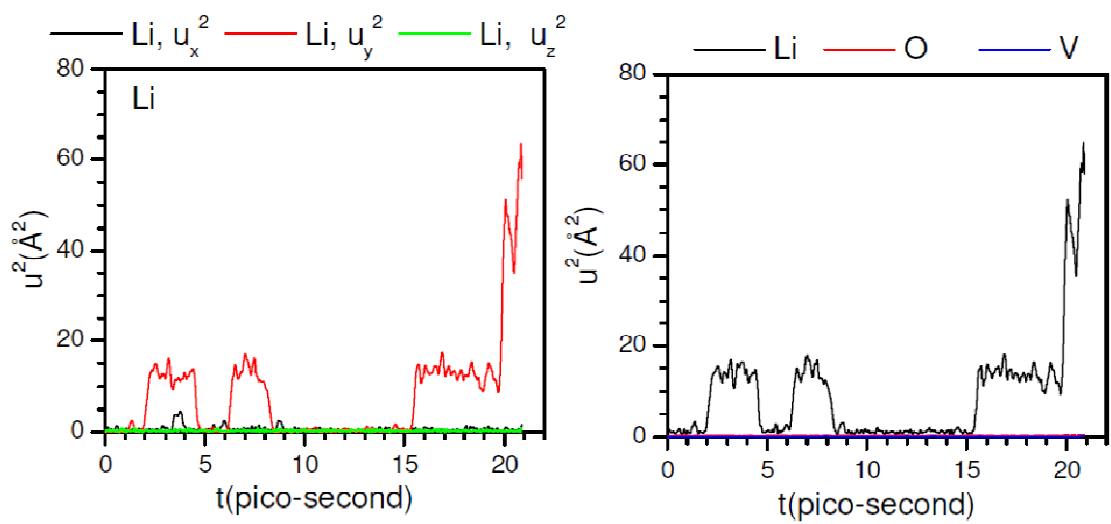

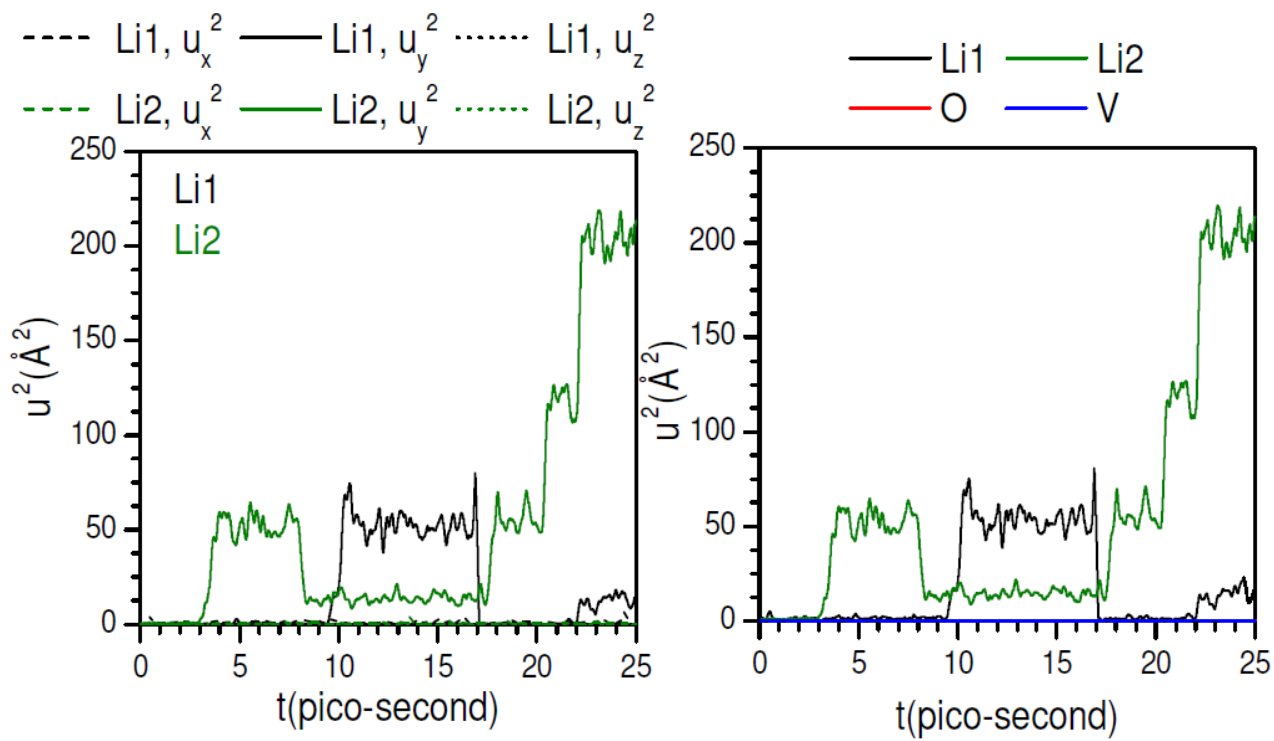

**FIG 4** (Left) The calculated squared displacement of Li atoms (Li1 & Li2) along the crystallographic axes at 1200K. (Right) The calculated mean squared displacement of various atoms Li1, Li2, O, Al in $Li_{0.16}V_2O_5$ obtained from ab-initio MD simulations at 1200 K.

**FIG 5** The calculated trajectories of Li in **(Left)** $Li_{0.08}V_2O_5$ and **(Right)** $Li_{0.16}V_2O_5$ obtained from ab-initio MD simulations at 1200 K. The time dependent positions of Li atoms are shown by green and black dots. Key: O- Red & V-Blue.

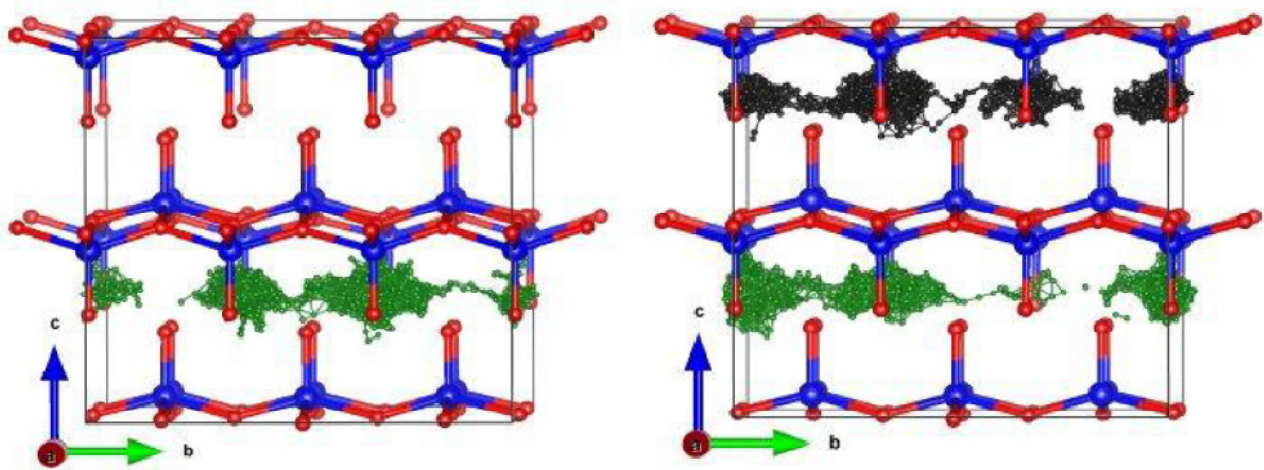